\documentclass[aip]{revtex4-1}

\usepackage{dcolumn,graphicx}
\usepackage{upgreek}
\usepackage{float}
\usepackage{natbib}
\usepackage{amsmath}
\usepackage[usenames,dvipsnames]{color}
\usepackage{soul}
\usepackage{pdfpages}

\begin{document}

\title{\Large On-Chip Detection of Entangled Photons by Scalable Integration of Single-Photon Detectors}

\author{Faraz Najafi}%
\thanks{These authors contributed equally to this work.}
\affiliation{Department of Electrical Engineering and Computer Science, Massachusetts Institute of Technology, 77 Massachusetts Avenue, Cambridge, MA 02139, USA}
 
\author{Jacob Mower}%
\thanks{These authors contributed equally to this work.} 
\affiliation{Department of Electrical Engineering and Computer Science, Massachusetts Institute of Technology, 77 Massachusetts Avenue, Cambridge, MA 02139, USA}

\author{Nicholas C. Harris}%
\affiliation{Department of Electrical Engineering and Computer Science, Massachusetts Institute of Technology, 77 Massachusetts Avenue, Cambridge, MA 02139, USA}

\author{Francesco Bellei}%
\affiliation{Department of Electrical Engineering and Computer Science, Massachusetts Institute of Technology, 77 Massachusetts Avenue, Cambridge, MA 02139, USA}

\author{Andrew Dane}%
\affiliation{Department of Electrical Engineering and Computer Science, Massachusetts Institute of Technology, 77 Massachusetts Avenue, Cambridge, MA 02139, USA}

\author{Catherine Lee}%
\affiliation{Department of Electrical Engineering and Computer Science, Massachusetts Institute of Technology, 77 Massachusetts Avenue, Cambridge, MA 02139, USA}

\author{Prashanta Kharel}%
\affiliation{Department of Electrical Engineering, Columbia University, 1300 SW Mudd, MC4712, 500 West 120th Street, New York, NY 10027, USA}

\author{Francesco Marsili}%
\affiliation{Jet Propulsion Laboratory, California Institute of Technology, 4800 Oak Grove Drive, Pasadena, California 91109, USA}

\author{Solomon Assefa}%
\affiliation{IBM TJ Watson Research Center, Yorktown Heights, NY 10598, USA}

\author{Karl K. Berggren}
\thanks{contact: berggren@mit.edu, englund@mit.edu}
\affiliation{Department of Electrical Engineering and Computer Science, Massachusetts Institute of Technology, 77 Massachusetts Avenue, Cambridge, MA 02139, USA}

\author{Dirk Englund}
\thanks{contact: berggren@mit.edu, englund@mit.edu}
\affiliation{Department of Electrical Engineering and Computer Science, Massachusetts Institute of Technology, 77 Massachusetts Avenue, Cambridge, MA 02139, USA}

\date{18 April 2014}

\maketitle

\begin{bf}

Photonic integrated circuits (PICs) have emerged as a scalable platform for complex quantum technologies using photonic and atomic systems~\cite{2012.NPhys.Walther.photonic_quantum_sim, obrien_phot_quant_tech_2009, 1367-2630-12-3-033031}. A central goal has been to integrate photon-resolving detectors to reduce optical losses, latency, and wiring complexity associated with off-chip detectors.  Superconducting nanowire single-photon detectors (SNSPDs \cite{2001.APL.Goltsman, 2009.Nphot.Hadfield}) are particularly attractive because of high detection efficiency \cite{2013.NPhoton.Marsili-Nam.93_perc_SNSPD}, sub-50-ps  timing jitter \citep{2009.JMO.Dauler-Berggren.photon_number_resolution_SNSPD}, nanosecond-scale reset time \cite{2011.apl.marsili_najafi}, and sensitivity from the visible to the mid-infrared spectrum~\cite{2012.NanoLett.Marsili.SNSPD}. However,  while single SNSPDs have been incorporated into individual waveguides~\cite{2011.APL.Fiore.WG_SNSPD, 2011.ArXiv.Tang.SNSPD}, the system efficiency of multiple SNSPDs in one photonic circuit has been limited below 0.2\%~\citep{2013.OptEx.Fiore.2-SNSPD-WG, zwiller2013} due to low device yield~\cite{2007.APL.Voronov.SNSPD-constriction}. Here we introduce a micrometer-scale flip-chip process that enables scalable integration of SNSPDs on a range of PICs. Ten low-jitter detectors were integrated on one PIC with 100\% device yield. With an average system efficiency beyond 10\% for multiple SNSPDs on one PIC, we demonstrate high-fidelity on-chip photon correlation measurements of nonclassical light. 

\end{bf}

\hspace{0.5in}

Photonic integrated circuits are being developed for a wide range of applications in quantum information science, including quantum simulation\citep{Spring15022013, Broome15022013, 2012.NPhys.Walther.photonic_quantum_sim, 2011.ACM.Aaronson.boson_sampling}, quantum photonic state generation \cite{2014.NatPhoton.quantum_sources, Chen:11,Fukuda:05,2011.PRA.Mower.AMPP}, quantum-limited detection~\cite{2009.PRA.Guha-Erkman.QI_receivers}, and linear optical quantum computing\cite{2008.Science.OBrian.quantum_circuit, obrien_phot_quant_tech_2009, Nielsen04PRL,KLM01}. These applications require multiple detectors with low timing jitter. The lowest timing jitter for infrared photon detection has been achieved with SNSPDs based on sub-100-nm-wide and $\sim$~4-nm-thick niobium nitride (NbN) nanowires. However, to date there has been no scalable approach to integration of SNSPDs into photonic circuits: while single isolated waveguide-integrated SNSPDs have been demonstrated~\cite{2011.APL.Fiore.WG_SNSPD, 2011.ArXiv.Tang.SNSPD}, the highest reported system detection efficiency for just two SNSPDs integrated into the same photonic circuit remains significantly below 1\% ~\citep{2013.OptEx.Fiore.2-SNSPD-WG, zwiller2013}. The central challenge when building systems with multiple SNSPDs remains the low fabrication yield, which is limited by defects at the nanoscale~\cite{2007.APL.Voronov.SNSPD-constriction}. This yield problem is exacerbated when such detectors are integrated onto photonic chips, which can require tens of additional fabrication steps of their own. 
Here we report on a micrometer-scale flip-chip process developed to overcome the yield problem by separating the PIC and the SNSPD fabrication processes. Our approach is compatible with a wide range of PICs, including CMOS-compatible silicon photonics, in a back-end-of-the-line step.

\begin{figure}[H]
\includegraphics[width=6.5in]{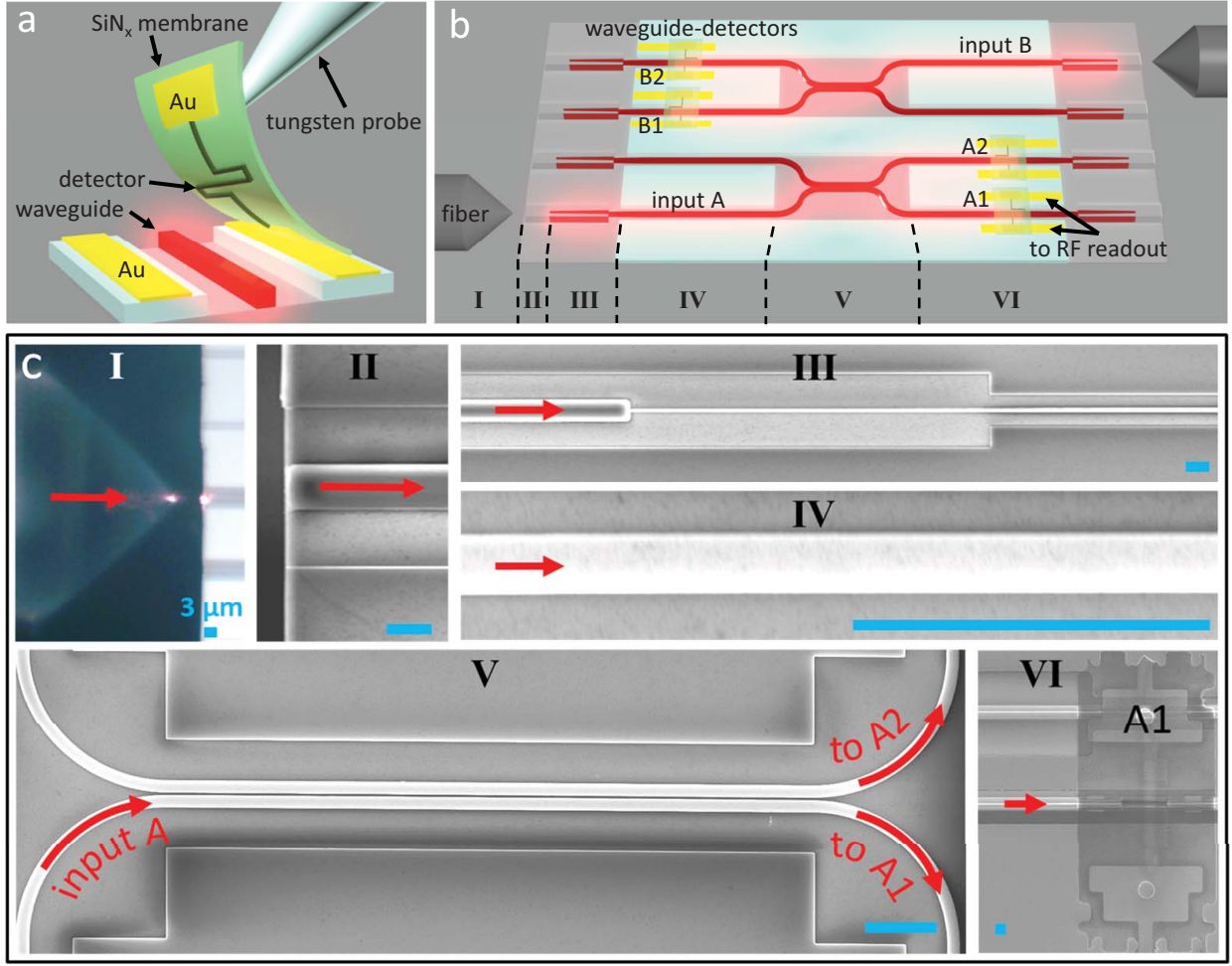}
\caption{{\small  (a) Membrane transfer of an SNSPD onto a photonic waveguide. (b) Sketch of photonic chip with four waveguide-integrated detectors (A1, A2, B1 and B2). (c) Micrographs of sections I-VI labeled in (b). Infrared light (red arrows) was coupled from a lensed fiber (I) with a spot diameter of 2.5~$\upmu$m into a 2~$\upmu$m $\times$ 3~$\upmu$m polymer coupler (II). The coupler overlapped with a 50- to 500-nm-wide inverse-tapered section of a silicon waveguide (III). The input light traveled along the $500$-nm-wide waveguide (IV) over a distance of 2 mm before reaching a 50:50 beamsplitter (directional coupler in V) followed by the waveguide-integrated detectors (VI). The equivalent length of the scale bar (blue) is 3 $\upmu$m. }}
\label{fig:figure1}
\end{figure}

Fig.~1(a) outlines the elements of the assembly process. Hairpin-shaped SNSPDs~\citep{2011.APL.Fiore.WG_SNSPD, 2011.ArXiv.Tang.SNSPD} were fabricated on $\sim$~200-nm-thick silicon nitride (SiN$_x$) membranes; silicon-on-oxide (SOI) PICs were fabricated separately (see Methods). After evaluating the SNSPDs in a cryostat, high-performance detectors were selected from the fabrication chip and transferred onto the desired SOI waveguides. Using this method, we assembled a proof-of-concept photonic circuit, shown in Fig.~\ref{fig:figure1}(b), comprising an optical network with two input and four output ports, each coupled to an SNSPD. We measured an estimated on-chip detection efficiency up to 45\% for 1550-nm-wavelength single photons and timing jitter as low as 42 ps. The light was coupled into the waveguides using inverse tapered couplers with $\sim$ 3 dB insertion loss \cite{2003.OpEx.Vlasov.WG-PC-integration}, resulting in a system detection efficiency (from the external fiber) up to $19\pm 2\%$.  This system efficiency enables the first on-chip intensity autocorrelation measurements of nonclassical light, demonstrated here for  photon pairs generated by spontaneous parametric down conversion. 
 
\newpage 

The detector comprised multiple nanowires connected in parallel (see SI), as shown in Fig.~2(a). This SNSPD variant~\cite{Ejrnaes1, 2011.NanoLett.Marsili-Berggren.SNSPD} has been shown to double the signal-to-noise ratio of the photodetection voltage compared to traditional single-wire SNSPDs. The detector length was designed using a finite-element model~\citep{2009.IEEE.Hu-Berggren.efficient_coupling_SNSPD} to ensure optical absorption exceeding 50\% (see SI). 

\begin{figure}[H]
\includegraphics[width=6.5in]{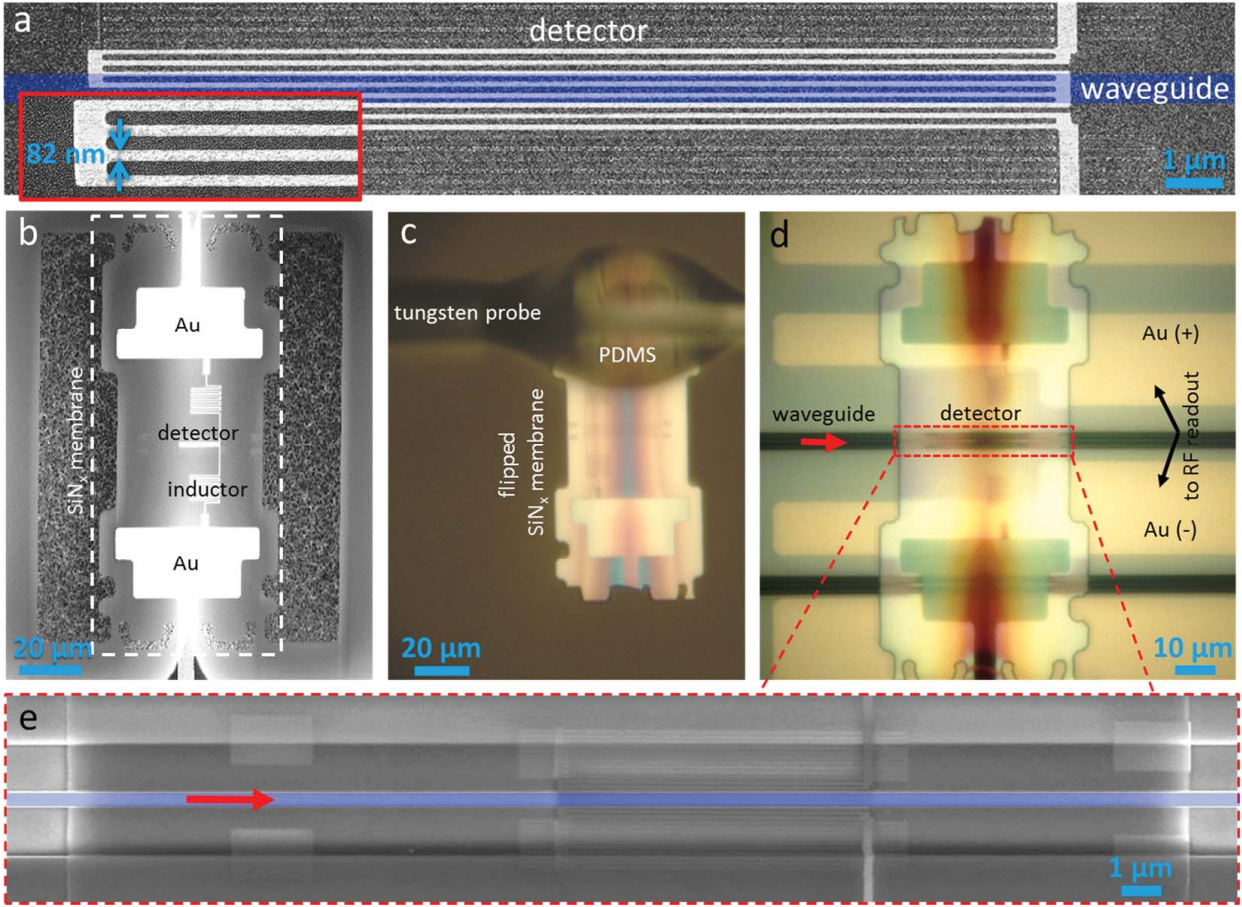}
\caption{{\small  (a) Scanning Electron Micrograph (SEM) of an SNSPD based on 82-nm-wide superconducting nanowires (see inset). The purple strip marks the intended location of the waveguide after the integration is complete. (b) SEM of suspended SiN$_x$ membrane with detector on top. The area of the membrane was 50 $\upmu $m $\times$ 120 $\upmu$m. (c) The detector was removed from the carrier chip using a tungsten microprobe containing a drop of hardened PDMS near the tip. The membrane was then flipped and the detector aligned to the waveguide under an optical microscope; this step simultaneously established electrical contact to Au strips on the photonic chip. (d) Optical micrograph of an SNSPD integrated with a Si waveguide. (e) SEM of waveguide-integrated detector in the region marked by a dashed line in (d). The silicon waveguide is highlighted in purple.  }}
\label{fig:figure2}
\end{figure}

We fabricated 225 detectors on a $\sim$~200-nm-thick SiN$_x$ layer over a Si substrate. The underlying silicon was then etched (see Methods), leaving hundreds of free-standing membranes carrying SNSPDs. One of these suspended membranes is shown in Fig.~\ref{fig:figure2}(b). Each membrane was connected to the bulk substrate through six narrow ($\sim$ 2-$\upmu$m-wide) bridges, two of which connected the detector on the membrane electrically to large contact pads on the bulk substrate for testing the detectors after the etch step (see SI).

We characterized all detectors to identify low-jitter, high-efficiency devices (typically about 30\% of the detectors). As shown in Fig.~\ref{fig:figure2}(c), we removed selected detector membranes from the substrate using tungsten microprobes coated with polydimethylsiloxane (PDMS) adhesive. We then placed  membranes detector-side-down onto the target waveguide with sub-1-$\upmu$m alignment accuracy under an optical microscope. For electrical readout, the gold pads on the membranes contacted complementary pads on the PIC (Fig.~\ref{fig:figure2}(d)). These gold-gold contacts withstood repeated thermal cycles with no noticeable degradation (see SI). Fig.~\ref{fig:figure2}(e) shows the resulting waveguide-integrated detector.  Because we transferred only high-performance detectors, we were able to achieve perfect yield in the assembled device, resolving the non-scalability of low-jitter SNSPD fabrication~\cite{2007.APL.Voronov.SNSPD-constriction}. 

\newpage

Using this process, we integrated four detectors (labeled A1, A2, B1 and B2) on a PIC and characterized the performance of the PIC shown in Figs. \ref{fig:figure1}(b,c) using four parameters: system detection efficiency (SDE), on-chip detection efficiency (ODE),  FWHM timing jitter (TJ), and noise-equivalent incident power (NEIP). The SDE includes all losses (i.e., coupling and transmission) between the fiber port outside the cryostat and the detector. We determined the SDE from the ratio of the SNSPD photocount rate to the photon flux coupled into the fiber port (see SI). Our chip reached an SDE of 19\% for input A (11\% for A1 and 8\% for A2) and 7\% for input B (3\% for B1 and 4\% B2). These SDE values represent an improvement of two orders of magnitude compared to previous approaches for multi-detector integration~\cite{2013.OptEx.Fiore.2-SNSPD-WG}. 

The ODE is defined as the probability that a photon already coupled into the waveguide is detected~\cite{2011.ArXiv.Tang.SNSPD, 2013.OptEx.Fiore.2-SNSPD-WG} (see SI). We estimated the ODE as SDE/$\eta_c$, where $\eta_c = 0.25$ accounts for coupling losses into the PIC (3 dB) and the splitting ratio of the directional couplers before the SNSDPs  (3 dB). The transferred detectors reached ODEs between 12\% and 45\% and 42- to 65-ps TJ. 

\begin{figure}[H]\centering
\includegraphics[width=6.5in]{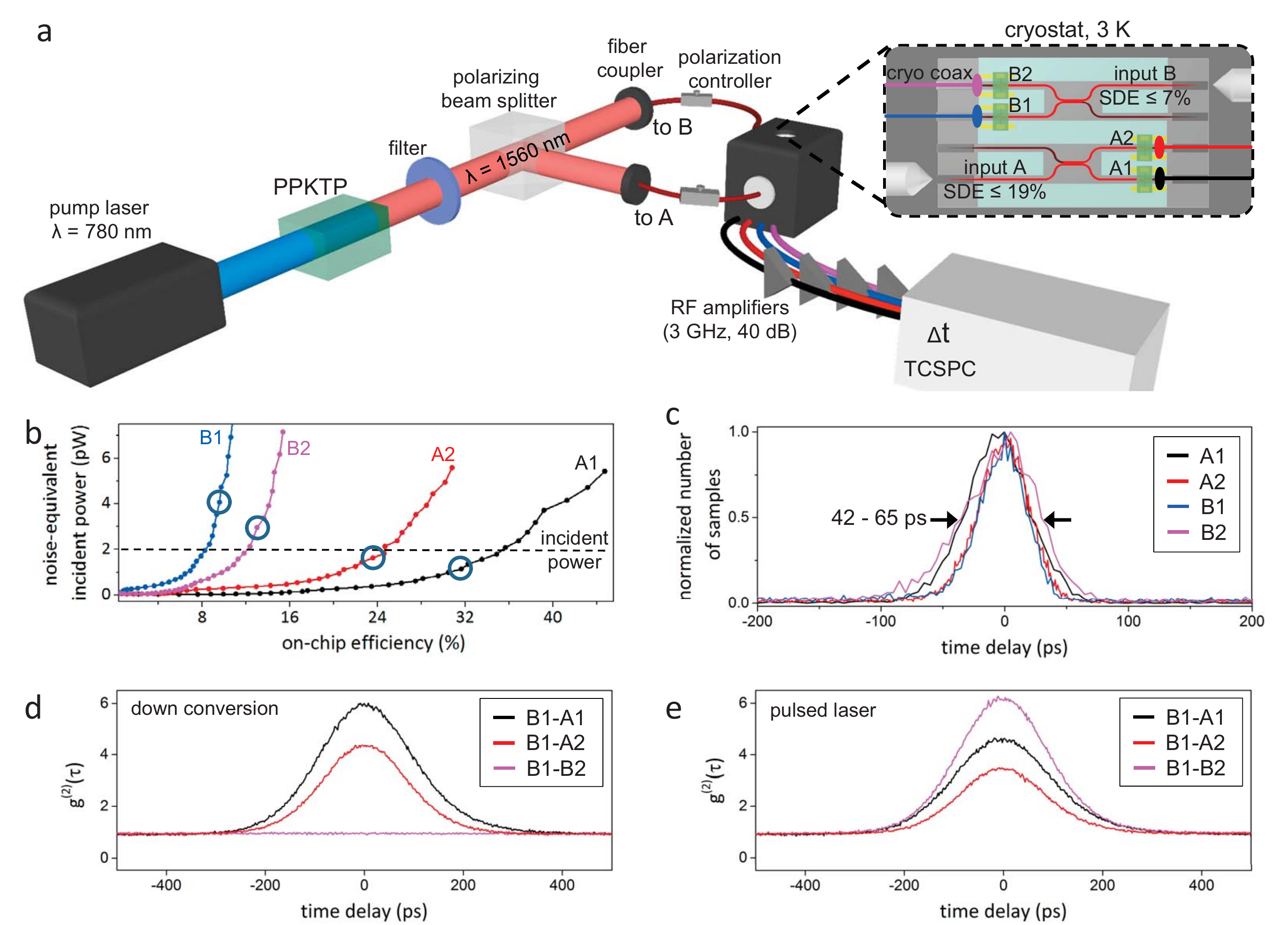}
\caption{{\small  (a) Experimental setup for on-chip $g^{(2)}_{AB}(\tau)$-measurements of an entangled-photon source coupled into the PIC (cooled to 3 K). (b) Noise-equivalent incident power vs. on-chip efficiency for the detectors shown in Fig.~\ref{fig:figure1}(b). The circles mark the operation points chosen for subsequent coincidence measurements. (c) Photodetection delay histogram of the detectors shown in Fig.~\ref{fig:figure1}(b) when operated at the maximum on-chip efficiency. (d, e) Coincidence counts vs. time delay between B1 and \{A1, A2, B2\} for the entangled-photon-pair source (d) and for a mode-locked sub-ps-pulsed laser (e). The average laser power was adjusted to match that of the photon-pair source.}} 
\label{fig:figure3}
\end{figure}

The NEIP is given by $ \mbox{SDCR} / \mbox{SDE} \cdot \hbar \omega$, where SDCR is the system dark count rate and $\hbar \omega =0.81$ eV. Fig.~\ref{fig:figure3}(b) shows the NEIP vs. ODE for the waveguide detectors on couplers A and B. The ratio of the power incident onto the detectors (IP) and the NEIP characterizes the signal-to-noise ratio for single-shot measurements. In this work, the NEIP was limited by radiation leakage (see SI) through a cryostat window used to image and align the lensed fibers to the polymer couplers (Fig. 1(c-I)). Hence, for subsequent measurements, we operated the detectors at lower ODEs of 10 - 32\% (circled points in Fig.~\ref{fig:figure3}(b)), which reduced the dark count rate and resulted in a ratio of IP/NEIP $\sim$ 0.5 - 1.7. 

We used these high-SDE SNSPDs to characterize time-energy entangled photon pairs entirely on the PIC. Entangled photon pairs were generated by spontaneous parametric down conversion (SPDC) from a 1-cm periodically poled potassium titanyl phosphate (PPKTP) waveguide, as shown in Fig.~\ref{fig:figure3}(a). Signal and idler photons of $\sim$ 1 ps duration and orthogonal polarization were separated using a polarizing beam splitter and sent into inputs A and B of the PIC. The SPDC pump power was adjusted to generate pairs at $\sim 1.5\cdot10^{8}$ Hz,  corresponding to a multi-pair probability of $\sim 4.4\cdot 10^{-4}$ per TJ. We obtained the second-order correlation function from $g^{(2)}_{AB}(\tau_i)=N_{AB}(\tau_i)/(r_Ar_B\Delta\tau_i T)$, where $N_{AB}(\tau_{i})$ is the measured number of coincidences between inputs A and B at time difference $\tau_{i}$, $r_{A}$ ($r_{B}$) is the count rate from input A (B), $\Delta\tau$ is the coincidence bin duration, and $T$ is the integration time. Fig.~\ref{fig:figure3}(d) shows the resulting $g^{(2)}_{AB}(\tau_i)$ function. Photon bunching is evident between inputs A and B, but not within individual channels (i.e., between A1 and A2 or B1 and B2), as expected for an entangled photon source.  The observed peak heights of $g^{(2)}_{AB}(0)\sim 4$ and $g^{(2)}_{AB}(0)\sim 6$ are lower than the theoretical value for ideal detectors  due to the finite  IP/NEIP ratio of our detectors (see Methods). By contrast, when  pulses from a mode-locked laser were injected into inputs A and B with average photon number per pulse greater than one, bunching was observed between all detector pairs (Fig.~\ref{fig:figure3}(e)), as expected for a pulsed classical source.

\newpage

\begin{figure}[H]\centering
\includegraphics[width=6.5in]{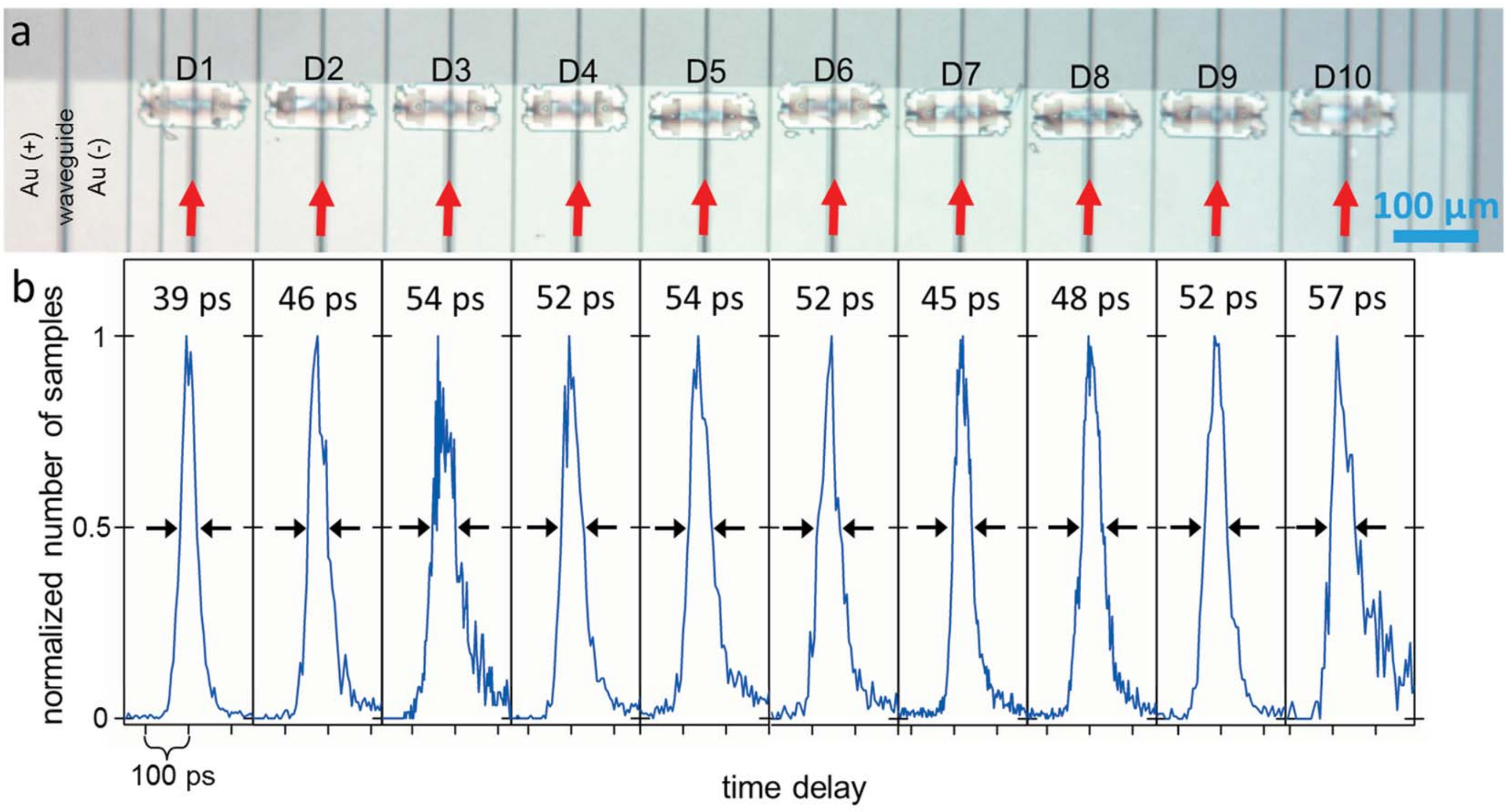}
\caption{{\small   (a) Optical micrograph of 10 waveguide-integrated detectors D1-10 assembled on the same photonic chip. The waveguides are marked by red arrows. (b) Top-illuminated photodetection delay histogram of D1-10 measured in a cryogenic probe station at 2.8 K base temperature. The timing jitter is listed above each histogram. }}
\label{fig:figure4}
\end{figure}

The ability to pre-select functioning devices enables scaling to more detectors with unity yield. Fig.~\ref{fig:figure4}(a) shows ten SNSPDs (D1-10) on adjacent waveguides with TJ values of 39 ps - 57 ps for 1550-nm-wavelength light. For rapid characterization, these devices were measured by top illumination in a cryogenic probe station. The photodetection delay histograms for all detectors are shown in Fig. 4(b).

The membrane transfer demonstrated here could be used to integrate other electro-optic devices, such as III-V lasers or single-photon sources, onto PICs. Since the device membrane is flexible, it conforms to the target chip, even if that chip is not perfectly flat. Because of the small size of the membrane, the process is also relatively tolerant to defects on the target chip, as opposed to processes involving large-area flip-chip bonding (e.g., see Ref.~\cite{Fang:06}), which require both surfaces to be free of defects. 

In conclusion, we have demonstrated the scalable integration of high-performance SNSPDs into photonic integrated circuits. We assembled ten adjacent waveguide-integrated detectors on a silicon PIC with 100\% yield and observed detector timing jitter values between 39 and 57 ps. Waveguide-integrated SNSPDs on the same PIC enabled on-chip $g^{(2)} (\tau)$-measurements of nonclassical light. Scaling to many tens to hundreds of detectors would ultimately be limited by the readout complexity. There is ongoing work to address this problem using electrical multiplexing schemes~\cite{zhao.apl.2013}. For more detectors, which require greater bandwidth, optical wavelength division multiplexing could be used, employing high-speed  ($>50$ GHz) modulators already available on PICs\cite{6086568}. The integration process demonstrated here is CMOS compatible; indeed, the PICs used in this experiment were fabricated in a CMOS compatible process with the exception of the polymer waveguide couplers, which can be replaced with SiN$_x$\cite{:/content/aip/journal/apl/55/23/10.1063/1.102290}. Thus, it appears likely that tens to hundreds of SNSPDs and other heterogeneous circuit elements can be integrated into high-performance PICs. This demonstration opens the door to fully integrated, high-performance photonic processors for quantum information science. 

\newpage

\section*{Methods}
\textbf{Detector fabrication.} A $\sim$~200-nm-thick SiN$_x$ layer was grown via plasma-enhanced chemical vapor deposition (PECVD) on double-polished silicon substrates. The NbN film was deposited on top of the SiN$_x$ layer via reactive magnetron sputtering (AJA system) at a substrate holder temperature of 800 $^{\circ}$C. The sheet resistance of the $\sim$ 4-nm-thick NbN films (thickness estimated from the deposition time) was 515~$\Omega$/square and the critical temperature was 10.9 K. Electrical contact pads were defined by UV-exposing a 700-nm-thick PMGI layer covered with 1.5-$\upmu$m-thick photoresist (S1813) for 13 seconds at 2300 $\upmu$W/cm$^2$ and developing the bilayer for 24 seconds in CD-26. This process achieved an undercut of the photoresist by $\sim$ 2 $\upmu$m, enabling smooth gold pad edges after liftoff. 10 nm Ti and 15 nm Au were evaporated and the liftoff was performed in acetone under sonication for 2 minutes followed by a 1-min dip in CD-26 and a 1-min DI dip. 70-nm-thick electron-beam-resist (HSQ) was spun on top of the sample, exposed in a 30 keV electron beam lithography tool (Raith 150, exposure dose 700-850 $\upmu$C/cm$^2$) and developed in TMAH at 27 $^{\circ}$C for 3 minutes. The HSQ pattern was transferred into NbN via a 2.5-min CF$_4$ reactive-ion etch (RIE) at 50~W. In order to improve electron-beam dose uniformity~\cite{yang2005fabrication}, additional features were exposed outside the hairpin-shaped detector. These dummy structures, also referred to as proximity-effect-correction features, are shown as parallel lines in dark grey outside the detector in Fig. 2(a).

\textbf{Detector suspension.} The detector was covered with S1813 and a trench pattern was exposed in the photoresist. This pattern was then used as an etch mask to define trenches around the detector through the SiN$_x$ layer via RIE with CF$_4$. This trench pattern left the underlying silicon substrate exposed. The silicon under the SiN$_x$ layer was removed using XeF$_2$, a selective isotropic etch gas. In the final step, the photoresist was removed in an NMP solution (see SI), resulting in a detector on a suspended SiN$_x$ membrane.

\textbf{PIC fabrication.} The PIC was fabricated on a 10 $\Omega$-cm, p-doped, 200-mm silicon-on-insulator (SOI) wafer from SOITEC. The wafer had a 220-nm-thick silicon device layer on top of a 2 $\upmu$m buried oxide layer. The 500-nm-wide silicon waveguides were fabricated on a CMOS line at the IBM Watson Research Center using electron-beam lithography. In a subsequent optical lithography step, SU8 polymer couplers were fabricated to allow sub-3-dB coupling loss from a lensed fiber to the silicon waveguide (see Ref. ~\cite{Vlasov:04} for further details). The gold pads on the PIC were fabricated in a similar manner to that outlined in the detector fabrication section above.

\textbf{Timing jitter measurements.} We used a mode-locked, sub-ps-pulse-width laser emitting at 1550 nm wavelength and 38 MHz repetition rate. The laser output was split into two SMF28 fibers, which we coupled to the detector under test and to a low-timing-jitter photodiode. The light coupled to the detector was attenuated to $<$ 5 pW and operation of the detector in single-photon regime was checked by confirming the linearity of the photocount rate as a function of incident photon flux (see SI). For detectors A1, A2, B1 and B2 the light was coupled to the waveguides A and B using a lensed fiber as shown in Fig.~\ref{fig:figure1}(b) and Fig.~\ref{fig:figure1}(c-I). The second sample, containing detectors D1-10, was back-illuminated with a high-NA fiber with light from the mode-locked laser, and single-photon operation regime was confirmed as described above. The electrical output from the detector and from the photodiode were sent to a 6-GHz-bandwidth, 40-GSamples/s oscilloscope.  We measured time delay $t_D$ between the detector pulse (start signal) and the pulse from the fast photodiode (stop signal). We acquired the instrument response function (IRF), a histogram of $>2000$ samples of $t_D$, and measured the timing jitter of the detector, which was defined as the FWHM of the IRF. 

\textbf{Correlation measurements.}
$g_{AB}^{(2)}(\tau)$ can be calculated from experimental data using the formula given in the main text. To incorporate detector dark counts, we define rates $r_{X}^{Y}$, where $X\in\{A,B\}$ (for channels $A$ and $B$, respectively) and $Y\in\{P,D\}$ (corresponding to a `photon' and `dark count,' respectively). $r_{A}^{D}$, for example, is the rate at which channel A registers dark counts, and $r_{A}\equiv r_A^D+r_A^P$ is the count rate on channel A. Now $g_{AB}^{(2)}(0)$ is
\begin{equation}
g_{AB}^{(2)}(0)=\frac{r_{A}^{P}\left(\eta_{H}+r_{B}^{D}\Delta\tau\right)+r_{A}^{D}\Delta\tau\cdot r_{B}}{r_{A}r_{B}\Delta\tau},
\end{equation}
where $\eta_H$ is the probability that channel B registers a photon given that channel A also registers a photon (i.e. the heralding efficiency) and $\Delta\tau$ is the bin duration. For $r_{A}^{Y}=r_{B}^{Y}\equiv r^Y$ and the ratio $K\equiv r^{P}/r^{D}$,
\begin{equation}
g_{AB}^{(2)}(0) = \left(\frac{K}{K+1}\right)^{2}\frac{\eta_{H}}{r^{P}\Delta\tau}+\frac{2K+1}{\left(K+1\right)^{2}}.
\end{equation}

In our experiment, $g_{AB}^{(2)}(0) \approx5$, which gives an estimate of the heralding efficiency, $\eta_H=3.5\cdot10^{-3}$.
\begin{acknowledgments}

This work was supported by DARPA Information in a Photon program, through grant W911NF-10-1-0416 from the Army Research Office, and the NSF (grant ECCS-0823778). F. Najafi was supported by the Claude E. Shannon Fellowship. J. Mower and A. Dane were supported by the iQuISE fellowship. D.E. was supported in part by an IBM Faculty Award. The authors thank J. Daley, M. Mondol, L. Li, K. Sunter, Y. Ivry, X. Hu, A. McCaughan, Q. Zhao, AttoCube and Montana Instruments for technical support. 

\end{acknowledgments}

\section*{Author Contributions}
F.N., J.M., S.A., K.B. and D.E. conceived and designed the experiments. F.N., F.B., J.M., F.M., D.E. and K.B. designed the detectors. F.N. and A.D. fabricated the detectors. J.M., S.A. and D.E. developed the waveguide chip. C.L. and J.M. developed the SPDC source. F.N. and J.M. performed the experiments. N.H., J.M., P.K. and D.E. developed the membrane transfer process. N.H. performed  the membrane transfer. F.N., J.M., N.H., K.B. and D.E. prepared the manuscript. 

\bibliographystyle{apsrev4-1}

\bibliography{references_BibDesk7_Dirk2}

\newpage



\section*{Supplementary material}

Supplementary Information: On-Chip Detection of Entangled Photons by Scalable Integration of Single-Photon Detectors

# Table of Contents

Supplementary Information: On-Chip Detection of Entangled Photons by Scalable Integration of Single-Photon Detectors

1. **Detector design**

1.1 **Optical simulations**

We used a finite-element model to calculate the length of the detector that would provide sufficient absorption of the mode travelling in the waveguide. The simulated geometry, superposed by the electromagnetic mode profile, is shown in Fig. S1 (a). The detector was separated from the surface of the waveguide by residual resist layers used to fabricate the detector and the waveguide. We estimated the thickness of the residual resist layer as 20 to 80 nm. The 2D finite element simulations, performed in COMSOL, were used to calculate the imaginary part of the effective mode index $n_i$. Following Ref. [1], we then calculated the optical absorption α in the detector as α = 1 - exp(-2 $n_i$ 2π $L_C$ / 1.55), where $L_C$ is the detector (coupling) length in μm. Based on the calculated absorption in the detector, shown in Fig. S1 (b), the chosen detector length of 17 μm would ensure an optical absorption of > 50 %. In practice, we measured optical absorption values of 62 – 74 %. The optical absorption could be further increased by increasing $L_C$ [2].

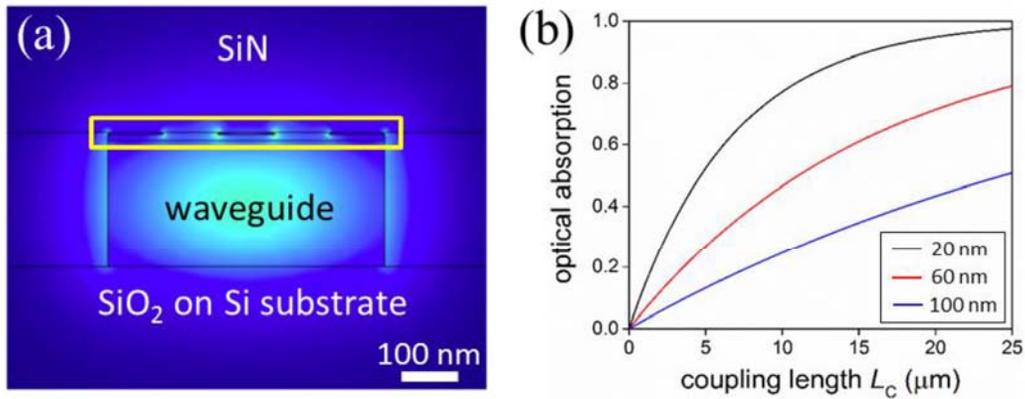

Fig. S1. (a) Cross-sectional geometry of waveguide-integrated detector (marked in yellow) superposed by the simulated spatial distribution of the intensity of the waveguide eigenmode. The detector consists of 80-nm-wide 4-nm-thick NbN nanowires arranged in a 200 nm pitch. The 500-nm-wide silicon waveguide was designed for 1550 nm center wavelength. (b) Calculated optical absorption in the detector vs coupling length for a residual resist thickness of 20, 60 and 100 nm.

1.2 **Nanowire circuit**

An SEM of the detector is shown in Fig. S2. The waveguide-detectors consisted of four units connected in series, with each unit comprising two ~80-nm-wide nanowires (200 nm pitch) in parallel. Detectors comprising this parallel-nanowire structure are commonly referred to as superconducting nanowire avalanche photodetectors (SNAPs [3, 4]). This detector design, illustrated in Figs. S2(c, d), was similar to detectors in Ref. [5]. The value of the series inductor $L_S$ was chosen as 50 nH so that the total inductance in series with a single parallel-nanowire unit was about 3- to 7-times the series inductance of a single nanowire (see Ref. [6]). The detailed inductance values are listed in Table 1.

Supplementary Information: On-Chip Detection of Entangled Photons by Scalable Integration of Single-Photon Detectors

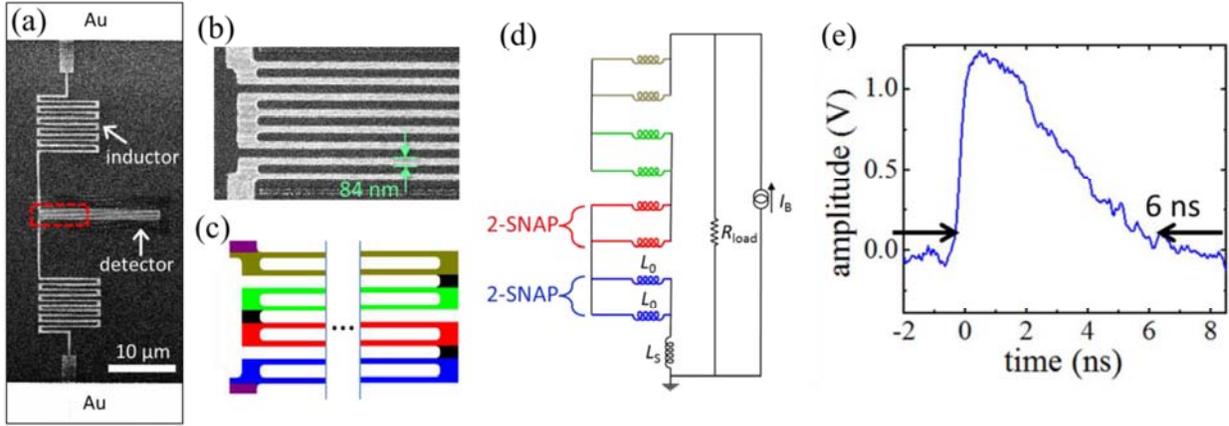

Fig. S2. (a) SEM of waveguide-detector. The detector is shown with the series inductor consisting of 300-nm-wide nanowires. (b) Magnified SEM of detector region encircled with red dashed lines in (a). The detector comprised four parallel-nanowire units in series. (c) Sketch representing the nanowire arrangement of the detector shown in (a, b). The detector consisted of four units in series, each comprising two parallel nanowires. (d) Equivalent circuit diagram for the detector. (e) Measured single-shot voltage trace of the output pulse of a detector with the geometry shown in (a)-(d).

| Nanowire width (nm) | Detector length (μm) | Nanowire length per SNAP section (μm) | #squares, $L_{kin}$ per section (nH) [assuming 80pH/square] | $L_S \geq$ 1.5 x $L_{kin}$ per section | Total inductance L of detector (nH) = $L_S+2*(L_{kin}$ per section) | Estimated reset time =3*L/50Ω (ns) |
|---|---|---|---|---|---|---|
| 80 | 17 | 17 | 213, 17 | 50 [2.9x] | 84 | 5 |

Table 1. Calculated inductance values for series-2-SNAPs based on 60- to 100-nm-wide nanowires. These values were used to design the detectors. For each 2-SNAP, we need > 3*$L_{kin}$ of a single section in series to ensure that the detectors have a broad avalanche regime of at least 20% of the switching current of an unconstricted SNAP (see Ref. [3] for more details). Since every 2-SNAP has already three 2-SNAPs in series (3*1/2* $L_{kin}$ of single section), we only need to add $L_S \geq 1.5*L_{kin}$ of a single section as series inductor.

## 2. Membrane fabrication

### 2.1 Membrane layout

Fig. S3(a) shows a basic design of a suspended membrane-detector, connected to the bulk substrate via four ~15-μm-long microbridges. These bridges had an unpredictable breaking pattern (Fig. S3(b)), resulting in fractured $SiN_x$ pieces that could fall in between the membrane and the PIC chip surface and prevent tight contact between the detector and the waveguide chip. In order to avoid residual $SiN_x$ pieces we modified the bridge design as shown in Fig. S3(c): the bridges were shorter (~3 μm long) with a ~0.8- to 1.5-μm-long constriction in the middle section of the bridge, resulting in a preferred breaking region marked by the dashed red lines. With this improved design most membranes could be removed from the bulk substrate (Fig. S3(d)) without substantial residual $SiN_x$ pieces.

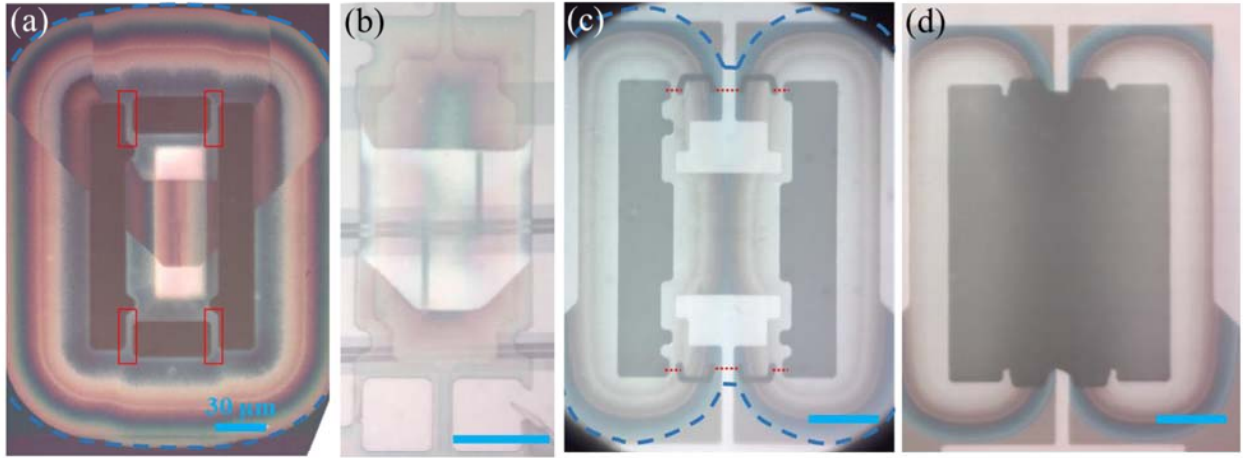

Fig. S3. Top-down optical micrographs of membrane-detectors. The scale bar in red represents 30μm in equivalent length. (a) Suspended membrane held with long microbridges (enclosed with red lines) and surrounded by four large trenches. The dashed blue lines separate the undercut SiN$_x$ region from the bulk susbstrate. (b) Transferred membrane with a design similar to the membrane shown in (a). (c) Suspended membrane with only two large trenches and short microbridges with constrictions. (d) Remaining structures on the primary SiN$_x$ chip after the membrane identical to the membrane shown in (c) has been removed.

## 2.2 Resist preparation and cleaning

During the membrane-detector fabrication process, illustrated in Fig. S4, photoresist layers covering the detector are used to define the outline of the membrane with trenches (Fig. S4 (b)) and to protect the detector during the Si etch step (Fig. S4 (c)). Initially the protective etch mask that was used to fabricate the trenches via reactive ion etch (RIE) with CF$_4$ was also used as a protective layer in the subsequent etch step with XeF$_2$. The fluorine gas (plasma) treatment during the RIE fluorinated the surface and hard-baked the resist, making it irremovable in solvents unless ultrasonic agitation was used. However, sonication could not be used after membrane undercut since it was found to cause membrane collapse. Oxygen-helium plasma (ashing) was the remaining option, but we could not remove the hard-baked residue after ashing, shown in Fig. S5(b).

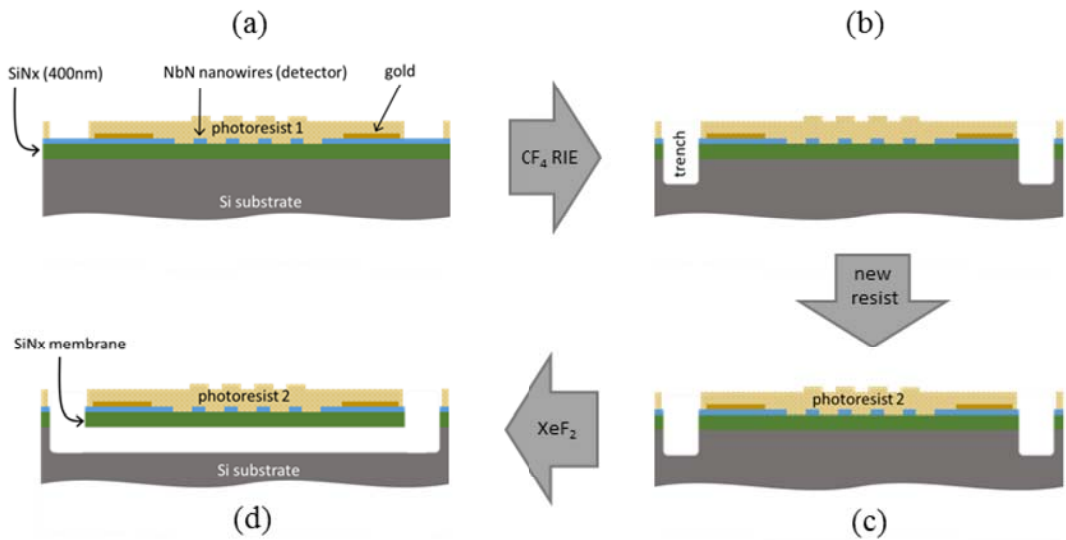

Fig. S4. Schematic cross-section illustrating the membrane-detector fabrication process.

Supplementary Information: On-Chip Detection of Entangled Photons by Scalable Integration of Single-Photon Detectors

We solved this issue by removing the resist mask after the trenches were fabricated via sonication ('resist 1' in Fig. S4(b)), and coating the detectors with a new resist mask for the silicon removal step ('resist 2' in Fig. S4(c)). Since the second mask was not exposed to a long $CF_4$ etch, we were able to remove it in an NMP-based resist stripper followed by an acetone and IPA rinse. While requiring an additional photolithography step, this stripping process did not leave a visible residue on the nanowires, as shown in Fig. S5(c).

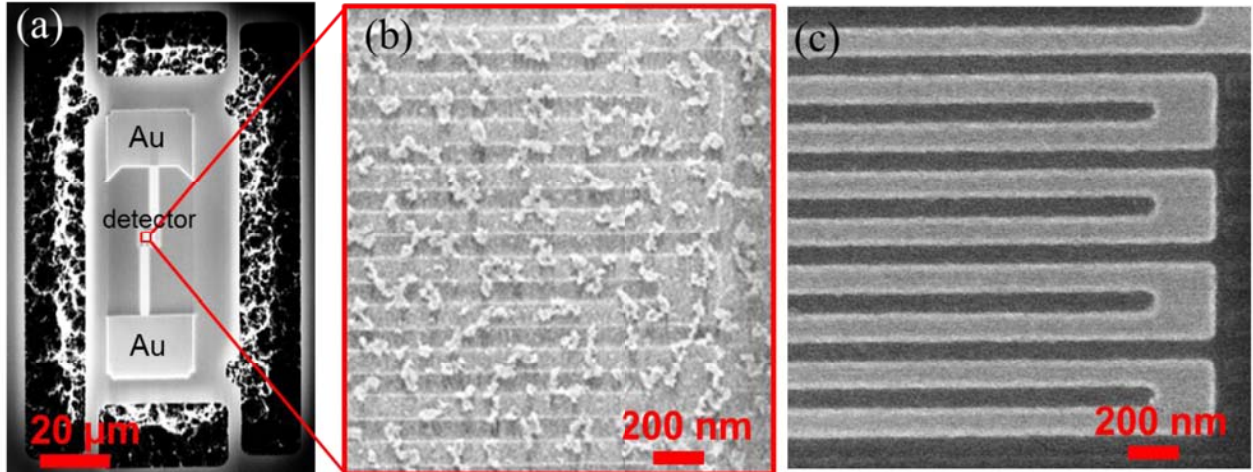

Fig. S5. (a, b) SEMs of membrane-detector after the protective photoresist was stripped in an oxygen plasma. (c) SEM of membrane-detector after the photoresist was stripped in an NMP solution.

## 3. Detector transfer

### 3.1 Transfer probe preparation

Polydimethylsiloxane (PDMS) was mixed in a 10:1 ratio with the curing agent and allowed to set for four hours. A tungsten microprobe (Ted Pella Autoprobe 100) was dipped in the PDMS solution, resulting in a PDMS droplet near the tip of the probe. The PDMS-covered probe was baked on a hot plate at 100 °C for 8 hours, followed by sonication in an ethanol-water mixture [7].

### 3.2 Membrane pickup and alignment accuracy

To remove the detector membranes from the substrate, three of the six microbridges connecting the membranes to the substrate (shown in Fig. 2(b)) were broken using a plain tungsten probe. The probe was then placed under the membrane and used to bend the membrane upwards, as shown in Fig. S6(b). A second tungsten probe, covered with PDMS droplet and mounted on a 6-axis micromanipulator, was then used to lift the membrane from the substrate, touching only the passive (back) side of the membrane (Fig. S6(c)). The PDMS served as an adhesive surface during the transfer (Fig. S6(d)) from the fabrication (carrier) chip to the PIC chip, where the membrane was then rotated, aligned and placed down under an optical microscope (Fig. S6(e)). After placement, the PDMS probe was used to press down on any regions of the membrane that exhibited interference fringes, indicating a separation between the PIC and membrane. Crucially, the detector surface was not in contact with any PDMS or other surfaces during membrane

Supplementary Information: On-Chip Detection of Entangled Photons by Scalable Integration of Single-Photon Detectors

pickup, minimizing contamination risk. Fig. S6(f) shows the SNSPD aligned to a silicon waveguide on a photonic chip using the highlighted alignment marks.

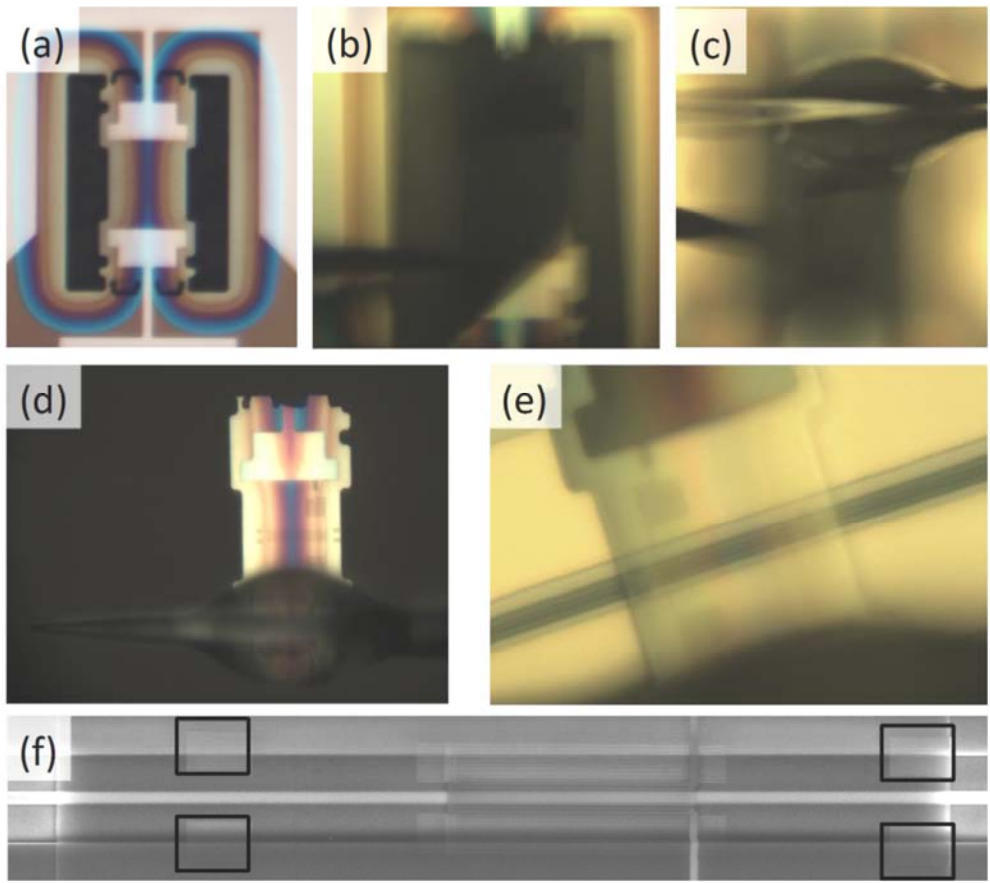

Fig S6. (a-e) Optical micrographs of SNSPD transfer process steps. (f) SEM of detector aligned to waveguide. Alignment marks, outlined in black, were used in the transfer process.

SEM images were taken in order to evaluate the alignment of the detectors with respect to the target waveguide. Fig. S7 shows the alignment marks, waveguide, and detector meander. The arrows in Fig. S7(a) reveal the boundaries between which the waveguide must exist for efficient detection. Of the four membrane-detectors placed, all contain detectors aligned to the waveguide.

Supplementary Information: On-Chip Detection of Entangled Photons by Scalable Integration of Single-Photon Detectors

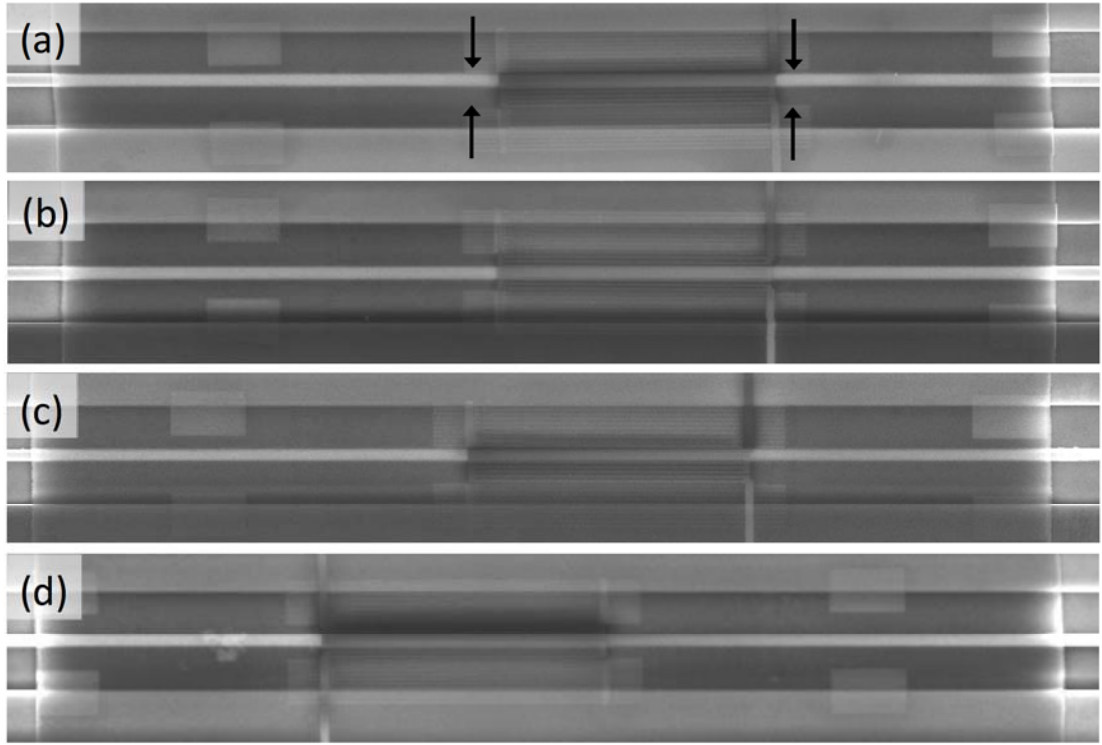

Fig. S7. SEM of four detectors (out of a total of four transfers) aligned to 500-nm-wide waveguides.

Efficient detection requires close contact between the SNSPD and the waveguide. Interference fringes serve as an indicator of the closeness of contact. The detector shown in figure S8(a) has little to no interference fringes visible, implying close contact between the membrane and waveguide chip surface. The opposite is true for the micrograph shown in figure S8(b). Here we can see visible fringing in the central region above the waveguide as well as near the gold pads. The detector shown in figure S8(b) would, in the best case, have poor detection efficiency and electrical properties. In the worst case it would exhibit no response to stimuli.

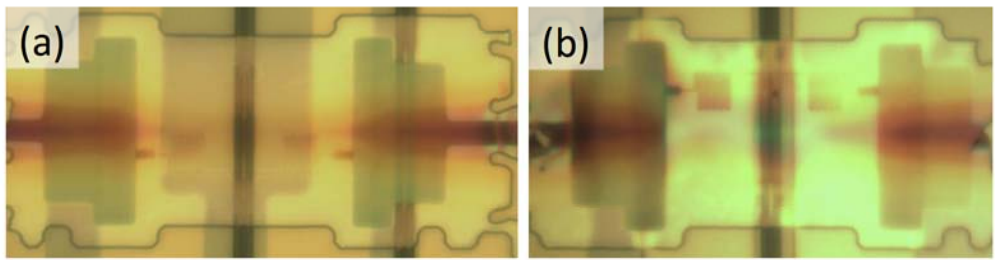

Fig. S8. Optical micrographs of two different detectors aligned to waveguides. (a) Membrane-detector with negligible interference fringes. (b) Membrane-detector with visible fringing.

Supplementary Information: On-Chip Detection of Entangled Photons by Scalable Integration of Single-Photon Detectors

## 3.3 Effect of membrane fabrication and thermal cycling on SNSPDs

Before performing membrane transfer, we characterized the room-temperature resistance $R_{after}$ of detectors suspended on membranes and compared to previous detector resistance values $R_{before}$ before the substrate was removed. Fig. S9(a) shows that the relative detector resistance change $(R_{after} - R_{before})/R_{before}$ was 1-2%, indicating no significant geometrical or material damage to the detectors due to the membrane fabrication process.

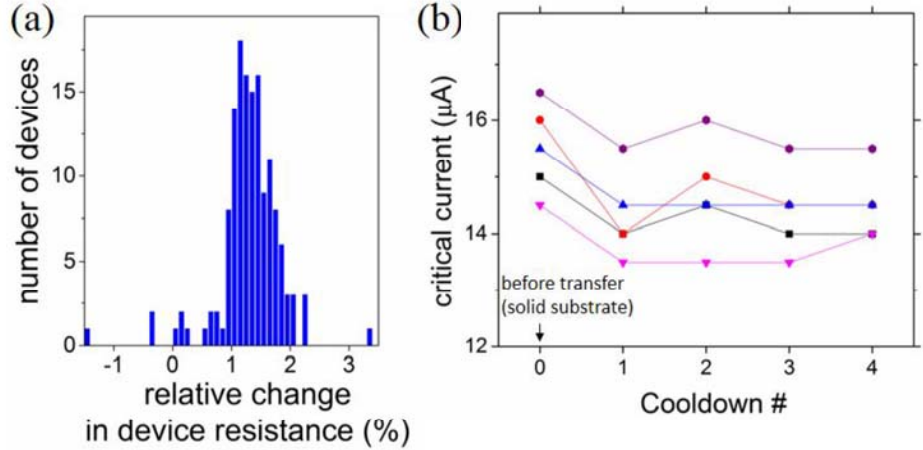

Fig. S9. (a) Histogram of relative change in SNSPD room-temperature resistance after undercut (shown in Figs. S4(c, d)) compared to the resistance values before undercut. (b) Critical current of detectors that were successfully transferred onto a secondary substrate on ~300-nm-thick $SiN_x$ membranes. Up to four thermal cycles were performed between ~2.8K and room temperature.

Fig. S9(b) shows the critical current of membrane-detectors that were successfully transferred onto a secondary substrate. The membranes consisted of ~300-nm-thick $SiN_x$. The critical currents of detectors on 300- to 400-nm-thick membranes were suppressed by ~10% compared to values measured on the solid substrate before undercut, while critical currents of detectors on sub-200-nm-thick membranes were suppressed by ~10-20%. Due to the small change in room temperature resistance values (Fig. S9(a)) we attribute the critical current suppression to the lower thermal capacity of the membranes compared to a solid substrate. Thermal cycling did not result in a measurable degradation (within the measurement accuracy of ~0.5 µA) of critical current of the transferred detectors.

Supplementary Information: On-Chip Detection of Entangled Photons by Scalable Integration of Single-Photon Detectors

## 4. Detector characterization

### 4.1 Optical absorption and critical current

After the detectors were transferred onto the photonic chip we measured the room-temperature optical absorption of the detector to confirm intimate detector-to-waveguide contact. The measured values were 74.3%, 73.9%, 65.3% and 62.1% for A1, A2, B1 and B2 respectively. The photonic chip was then mounted into a closed-cycle cryostat and the detectors operated at 3K base temperature. The critical currents after detector undercut and transfer (15.2 µA, 16.8 µA, 16.4 µA, and 14.8 µA) were ~ 20% lower in the critical current compared to pre-undercut values measured on the solid silicon substrate, possibly arising from the small thermal capacitance of the membranes.

### 4.2 Dependence of system dark count rate on shielding conditions

We used a closed-cycle cryostat with optical access to operate the chip shown in Fig. 1(b). The schematic cross-section of the cryostat is shown in Fig. S10(a). The PIC chip and micro-manipulated lensed fibers were kept at 3 K base temperature. In order to couple light from the lensed fibers into the waveguides, the edges of the chip, containing the polymer couplers (Figs. 1(c-I, c-II)), were imaged through the windows using a 50x long-working-distance objective. The direct imaging greatly simplified pre-alignment, while finer fiber-to-coupler alignment was performed using feedback from the SNSPDs. However, the optical access ports in this prototyping setup resulted in radiation leakage and therefore increased the system dark count rate of the detectors significantly, as shown in Fig. S10(b). When we replaced the 30 K window in the cryostat with a solid copper plate, we observed a significantly lower dark count rate of ~ 5 kcps instead of ~ 800 kcps at the operation point.

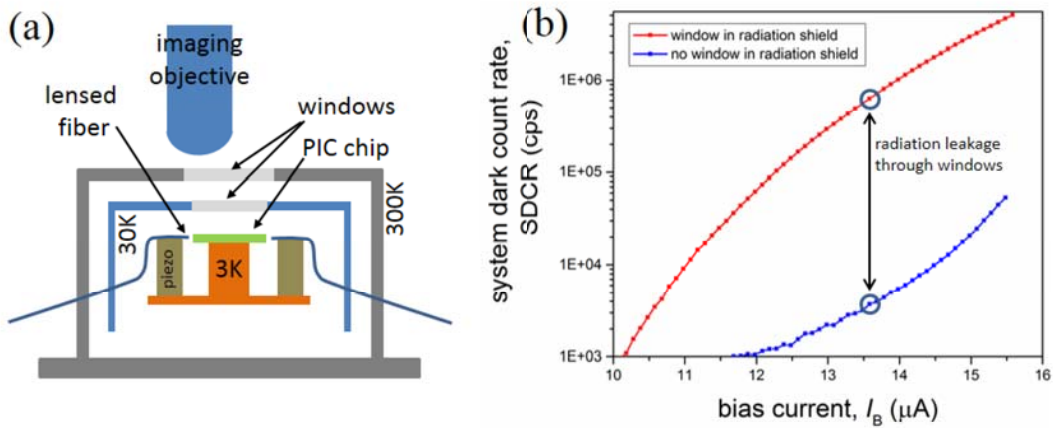

Fig. S10. (a) Schematic cross-section of cryostat used to operate the PIC chip. (b) System dark count rate (SDCR) curves representative of waveguide-integrated detectors operated in the cryostat shown in (a). The red curve shows the SDCR during the regular operation of the cryostat with windows, and the blue curve shows the SDCR with the windows replaced with copper plates.

Supplementary Information: On-Chip Detection of Entangled Photons by Scalable Integration of Single-Photon Detectors

## 4.3 Detection efficiency measurements

A schematic depiction of the experimental setup used to measure the system detection efficiency of the waveguide-integrated SNSPDs is shown in Fig. S11. Light from a fiber-coupled CW laser (Thorlabs S3FC1550, emitting at λ = 1550 nm, output power ~ 1mW) was split into two outputs. One output, used to monitor the power directly, was coupled to an InGaAs Photodiode (Thorlabs S154C), calibrated with a NIST-traceable curve down to 100 pW input power. The second output passed through a variable attenuator (JDS Uniphase HA9) and a polarization controller and coupled to the cryostat through a SMF28 fiber feedthrough. The calibration of the HA9 beyond the sensitivity of the photodiode was confirmed as follows: we recorded the SNSPD count rate under a given HA9 attenuation value (typically 50 - 80 dB), then replaced the HA9 with fixed fiber optic attenuators of the same attenuation value (Thorlabs FA series attenuators, separately calibrated at high laser power) and measured the SNSPD count rate again. Both count rates were generally within ± 10% of each other. Furthermore, we confirmed that the detector operated in single-photon regime during the system efficiency measurements, as demonstrated by the linearity of the photodetection count rate vs incident photon flux shown in Fig. S12(b).

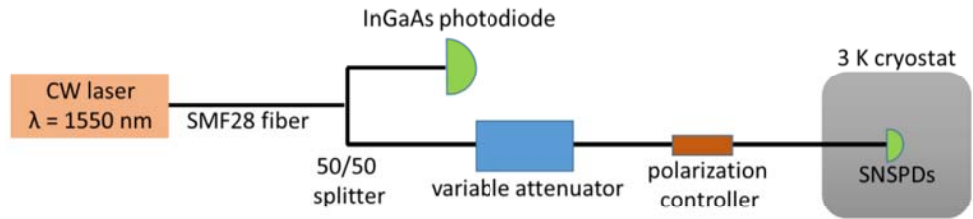

Fig. S11: Schematic depiction of experimental setup used to measure the system detection efficiency of waveguide-integrated SNSPDs.

The measured SDE for detectors A1, A2, B1 and B2 is shown in Fig. S12(a). We extracted the on-chip detection efficiency (ODE) the SDE as ODE=2 * 2 * (SDE per detector), therefore excluding ~ 3 dB fiber-to-waveguide coupling loss (comprising 2.7 ± 0.3 dB mode coupling loss and ~ 0.7 dB transmission loss in the lensed fiber) and the 3 dB splitting ratio of the directional coupler. All other on-chip losses, e.g. a ~ 1 dB transmission loss in the waveguide (3 dB/cm), are included in the ODE number.

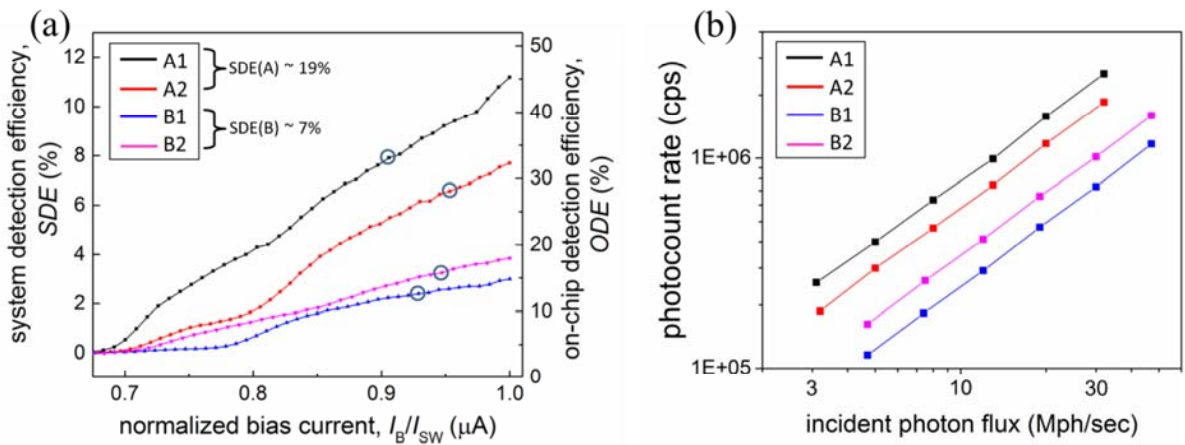

Fig. S12: (a) System detection efficiency vs. normalized bias current of the waveguide-integrated detectors shown in Fig. 1. The bias current ($I_B$) on the horizontal axis was normalized by the maximum bias current (switching current $I_{SW}$) of the detector. (b) Photocount rate in counts per second vs. incident photon flux for the detectors A1, A2, B1 and B2. The detectors were biased at the operation point marked by circles in (a). For the measurements shown in Figs. 3(b-e) the average photon flux was kept at ~10-15 million photons per second, which was well within the single-photon regime of the detectors.

Supplementary Information: On-Chip Detection of Entangled Photons by Scalable Integration of Single-Photon Detectors

## 5. Photon pair generation

We used a PPKTP waveguide source to generate entangled photons at 1560 nm. A 20 mW pump beam at 780 nm was focused on a PPKTP waveguide with cross section 2 μm × 4 μm. The waveguide was defined by ion implantation, and was 1 cm long. The phase matching bandwidth was approximately 1.5 nm, and the generated photon pair flux was estimated to be $1.5 * 10^8$ pairs/s. The down-converted signal and idler photons were coupled into a single fiber and split with a fiber polarizing beam splitter. The output fibers were coupled to polarization controllers, which were connected in turn to the fibers leading into the cryostat.


\end{document}